# Nagios Based Enhanced IT Management System

Ahmed D. KORA
Ecole Supérieure Multinationale des Télécommunications (ESMT), BP 10000
Dakar, Senegal

Moussa Moindze SOIDRIDINE
Department of Mathematics and Informatics, Université Cheikh Anta Diop,
Dakar, Senegal

**Abstract**: The integrated management of multi-providers equipments is a key asset for telecommunication operators or service providers when selecting the appropriate network and service management platform for their network. In this paper, we present an open and adaptable platform that support fault and configuration management for next-generation networks. This platform Named Nagios based – information technology management system (NB-ITMS) leverage of the well-known Nagios platform to implement new capabilities. Offering additional features applications and management functions enable easy and low cost management of advanced services and networks technologies. The performance of our platform has been compared to existing off-the-shell platforms.

1. INTRODUCTION

In the last decades, communications technologies have increasingly undergone a revolution. The fast emergence of new network technologies and the development of more and more heterogeneous equipments have resulted in a situation characterized by a high level of heterogeneity as well as complexity of centralized management solution because of the heterogeneity of underlying equipments to manage. The difficulty comes from the fact that the equipments are usually running proprietary management protocols and implementing heterogeneous management data models. In this complex environment, Information Technology (IT) and telecommunications networks operators are facing a crucial problem of management cost. Operators are therefore required to deploy various management platforms in order to manage the whole network. This situation will not change in the future as operator need to purchase equipments from manufacturers to reduce the risk of dependency from one manufacturer and also to reduce the cost of purchasing these equipments in a competing market. Consequently, the operator is forced to deploy specific management and monitoring tools for each line of products.  Nowadays, several supervision tools are able to use expert system engine to improve its management. However, the problem with such system is their limitations in term of reasoning in situation where new types of components are introduced in the network. In this case, most of the time the expert system efficiency to find an appropriate solution is not guaranteed. Other concepts of network management such as neuronal or alarm correlation have been also investigated in the literature but the new challenge is to be at least as effective as existing approaches and providing higher flexibility. A real telecom network is usually composed of at least a dozen of technical centers (TC) and switches (S), hierarchically structured. This network is managed from a supervision center (SC) which main tasks are operating and maintenance. Each component can receive or transmit messages. The problem at the monitoring system is its capacity to handle the flood of alarms sent by the different equipments and services called network elements.

 There are many challenges related to the tasks of acquisition, design, systems development and management [David Padi, (2009)]. This becomes progressively more difficult, requiring network managers to address the interoperability issues between heterogeneous management systems. The multitude of tools that have to be purchased after the integration of additional equipments in the network constitutes an important additional cost and a burden for the real business opportunities with the different manufacturers (Alcatel, Ericsson, Huawei,

Redline, Mikrotik ...). Key parameters considered in the choice of the network management system are the sizes of network elements and area, the diversity of components, services or applications supported by the network and management protocols required. The network topology and data communication network might impact the management efficiency. The different existing network elements management interfaces communication protocol could be classified into three. These protocols are CMIP (common management information protocol) introduced by ITU as telecommunication management solution, SNMP (simple network management protocol) introduced by IETF and then proprietary protocols. SNMP is the communication protocol used by Nagios. It is important to highlight that CMIP is a disappearing protocol because of the deployment of IP protocol. So, our choice for SNMP is motivated by this key parameter.

In the current context of competition, there is no management solution which could fit all situations. Networks operators, willing to deploy a centralized integrated management system, need to solve the difficult exercise of selecting the most appropriate tools to build their system. The available open and free solutions do not cover all the different management functions even those essential for efficient operation and maintenance. Finally proprietary solutions often expensive with limited network map view are adopted. The constraint of theses proprietary solutions is that they require specific and costly training. To overcome this problem of multiple management tools deployed in the same room in the case of centralized management approach, we have designed a new platform called NB-ITMS (Nagios Based - IT Management System). Nagios is an on open source supervision software easy to configure, free of charge. NB-ITMS has two more very important and critical enhanced features compare to Nagios: configuration management and enhanced GUI for operation and maintenance.

The remaining of this paper is organized as follows. In section 2, we describe our proposed NB-ITMS and its advantages. Section 3 is dedicated to the evaluation of the efficiency of NB-ITMS compared to other existing management supervision tools for fault and configuration management. Section 4, presents finally the conclusion of this work.

2. **NB-ITMS system description**

Nowadays, the budgets of IT departments are shrinking and license management is becoming more restrictive. IT departments require sophisticated tools to manage all the network equipments and services in the company in order to fulfill the requested quality of service (QoS) of the users and help them to compete efficiently in the nowadays fierce competing market.

A management system role is to interact with underlying network equipments and services in order to verify whether the services are provided with the required quality of service. Information needs to be is exchanged between equipments and services and the management system for this purpose. The global system could be conceptually represented by two blocs associated by inputs and outputs as depicted in Figure 1. These blocs named, as M and N, represent the management system and the managed resources respectively. $Y_M(t)$ corresponds to an external event or command initiated by an administrator or a technician in order to perform a particular management on the network modifying some network equipments or services parameters according to his observations.

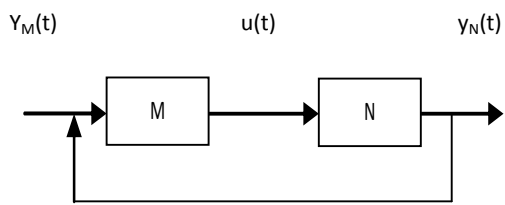

Figure 1: Global management system architecture

These modifications requests are executed by the management system and transmitted to the target equipment or service through a message u(t). According to that modification or any other event (alarm for example) raised in the network, the management system monitors the new network state denoted by $y_N(t)$ thanks to the management information base (MIB).

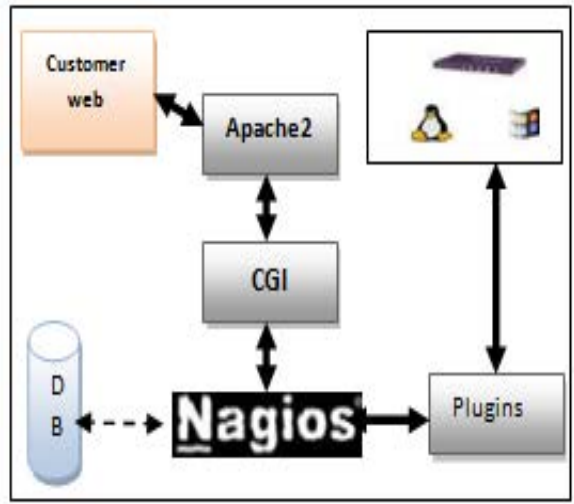

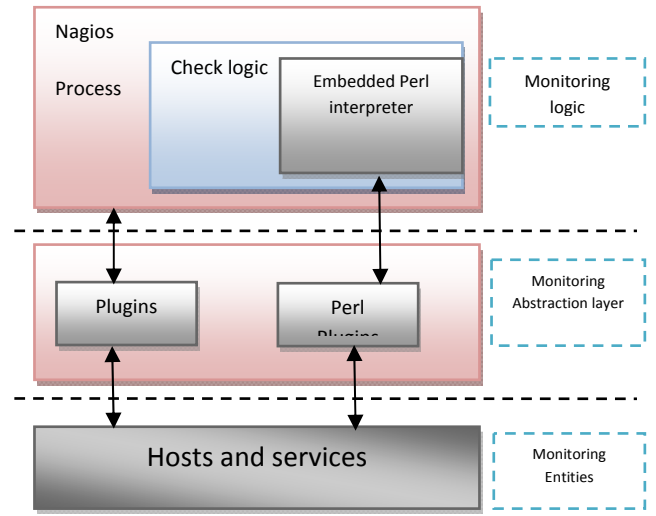

Figure 2: Default Architecture of Nagios

Figure 3: Layered architecture of Nagios

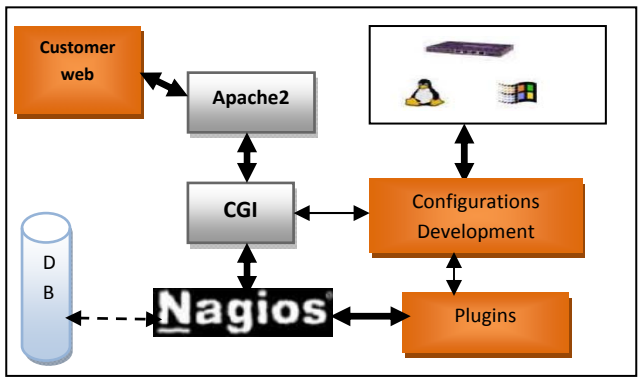

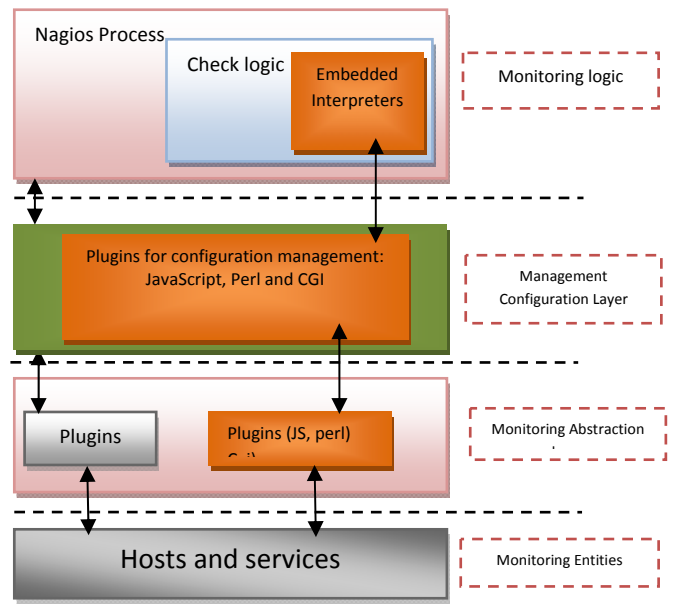

Figure 4: Architecture of NB-ITMS

Figure 5: Layered architecture of NB-ITMS

In this context, Nagios software is very popular during the last decade. Nagios is an open source supervision that respects the KISS (Keep It Simple, Stupid) principle of UNIX. Nagios relies on external programs commonly named plugins (Fig. 2). It is a modular program that facilitates the evolution of the system. Nagios is usually deployed as network monitoring software of network equipments and applications. It monitors selected hosts including servers and generates alarms whenever something wrong happens and controversy something good happens. Unlike many other monitoring tools, Nagios does not include any internal mechanisms for checking the status of equipments and services in the network. Nagios architecture described in Fig. 3 is composed mainly of three levels. The top level is monitoring logic (process core), the medium level is monitoring abstraction layer (plugins) and the bottom one is monitoring entities as user's web browser. The process logic core and embedded interpreters constitute the monitoring logic. The monitoring abstraction layer is the layer of adding existing plugins or newly developed. Plugins are compiled as executables or scripts (Perl scripts, shell scripts, etc.) that can be run from a command line to check the status of a host or a service. Nagios has to exchange information with the plugins to collect the results from them to determine the current status of hosts and services in the network.

One of the main limitations of Nagios is its inability to manage the configuration of underlying equipments. NB-ITMS which architecture is depicted by Fig. 4 is proposed to overcome this limitation. The plugins of the abstraction layer between the monitoring logic layer presented in the Nagios daemon and the actual services and hosts (Fig. 3) are dedicated for monitoring. There are already a lot of plugins that have been created in order to monitor basic resources such as processor load, disk usage, etc. There is no configuration plugins by default. Our approach has been to add new plugins for the configuration of managed devices. This concept leads to a new architecture denoted NB-ITMS already presented in Fig. 4 which has pointed out the additional configuration management bloc. The equivalent layered architecture is illustrated by Fig 5.

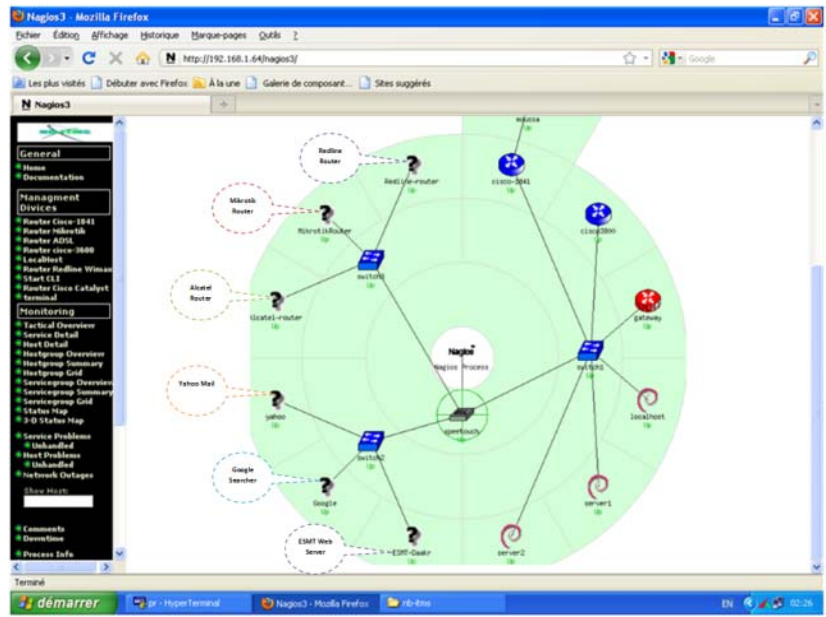

Figure 6: Nagios default setup MAP without personalize logos

Beside this, in the Nagios default configuration, all newly added components are not well represented in the MAP (graphic interface) (i.e. represented by an interrogation mark (?) because nagios is not capable to find out the appropriate image for them).

This was another limitation for the network administrator because he is not able to distinguish between these equipments on the map (Fig 6).

Ahmed D. Kora et al. / International Journal of Engineering Science and Technology (IJEST)

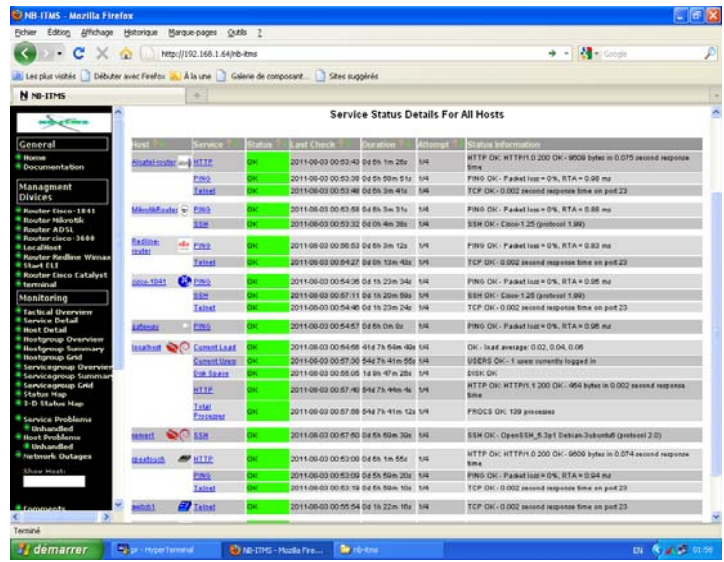

Figure 7(a): Services details after inserting MIBs

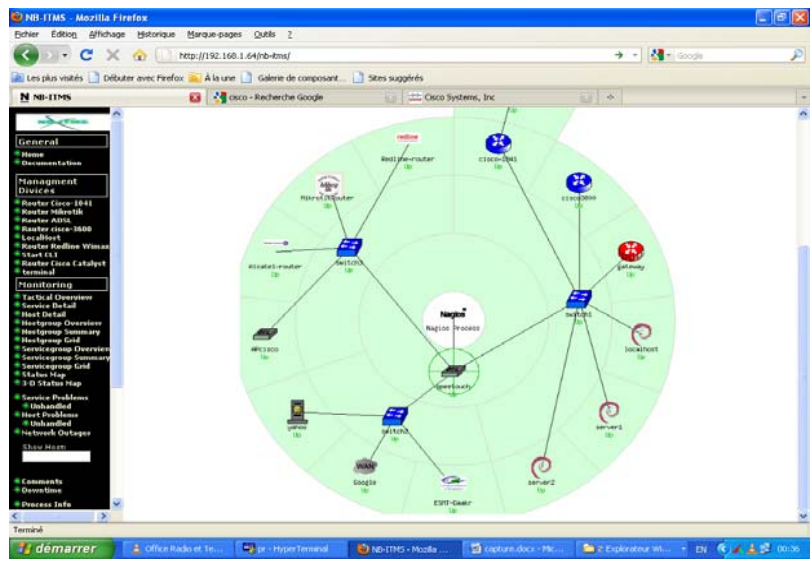

Figure 7(b): NB-ITMS setup MAP with logos

We have solved this problem in NB-ITMS developing a new mechanism that is capable to efficiently match between the equipment type and the corresponding logo (Fig.7). The most important challenge of NB-ITMS, compared to other open and free existing tools, has been the integration of management configuration function in the platform to enable directly remote configuration of network equipments using either web protocol or CLI (Command Line Interface). To achieve that, we have developed a new plugin that was integrated as a script in the platform.

3. **Evaluation of NB-ITMS Efficiency**

The evaluation of our management system efficiency requires choosing appropriate metrics. Different metrics such as CPU load, cost of communication between nodes (traffic), processor performance, delay, memory and network bandwidth, have been already considered in previous works [CHEN T. M.,(2002)].
Many state of art works have focused on the scalability of management systems by assessing their performance in face of a large number of management agents and metrics [Abdelkader Lahmadi, (2006)]. The authors of [Jogalekar, P., (2000)] have defined the performance as productivity of a system in term of relation between the rate of providing valuable services, the quality of those services and the cost of providing those services. Another definition is proposed in [Burness, A, (1999)], where performance is viewed as the response time, as seen by a user under normal working conditions, coupled with the cost of the system hardware requirements per user. We can also find several works related to the performance evaluation of management systems. Most of these studies have focused on the model of performance for management architectures and their associated cost. These performance models have quantified the response time of agents [Pattinson, C, (2001)], the volume of management traffic [Neisse, R.,(2004)] and resources usage [Subramanian, R.,(2000), Pras, A, (2004)]. However performance studies related to management architectures consider elementary performance metrics such as management requests rate and resource usage that will benefit from the management impact metric [11]. According to [Neisse, R., (2004)], we have proposed to use three metrics to evaluate the performance of our system. Those metrics are the quality (QoS) (i), the operationality (ii) and the cost (iii). The first (i) relates to the quality of operations based on the considered algorithm of management. The second metric (ii) combines all operations of management protocol such as management functions, deadlines, frequency of examination of the attributes and their rates of loss. The last metric (iii) combines the cost of management system and cost of all other activities on the managed network system.

All these considerations could be combined into a unique performance function P(k). Since, the performance function is intended to assess the feat variability of a management system. We have denoted by C(k) the function related to a metric of cost. A second function O(k) is a metric which quantifies the operationally aspect. The quality metric is represented by the function Q(k). Hence, the effectiveness of the management system at an impact factor k is given by :

$$P(k) = Q(k)*O(k)/C(k) \qquad (1)$$

Where the variable O(k) corresponds to the number of operations exchanged per unit of times between the management system and the managed nodes.

- the function Q(k) quantifies the quality of the work achieved under the factor considered, and

- the C(k) function is a metric of cost which includes the usage of the ressources ( processor, memory and bandwidth) per unit of time.

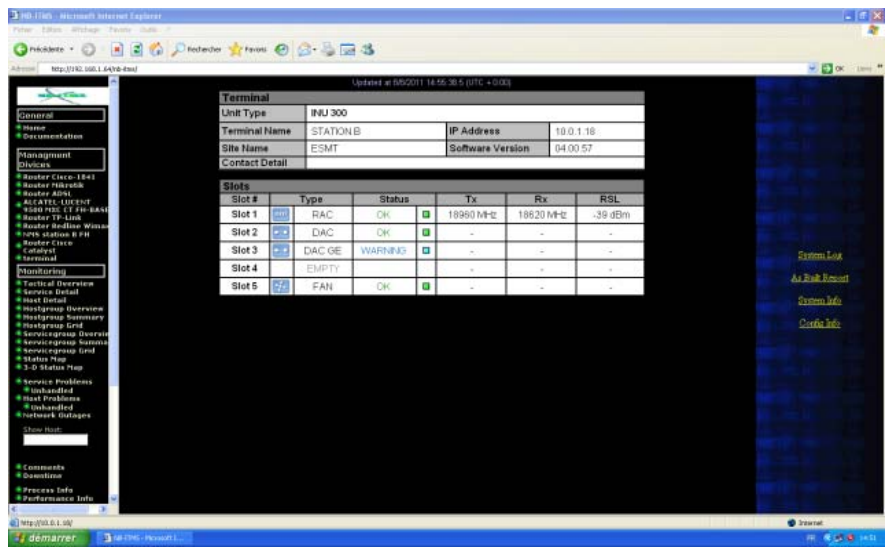

Figure 8a : View of supervision interface of Alcatel-Lucent microwave station named MXC 9500 via NB-ITMS

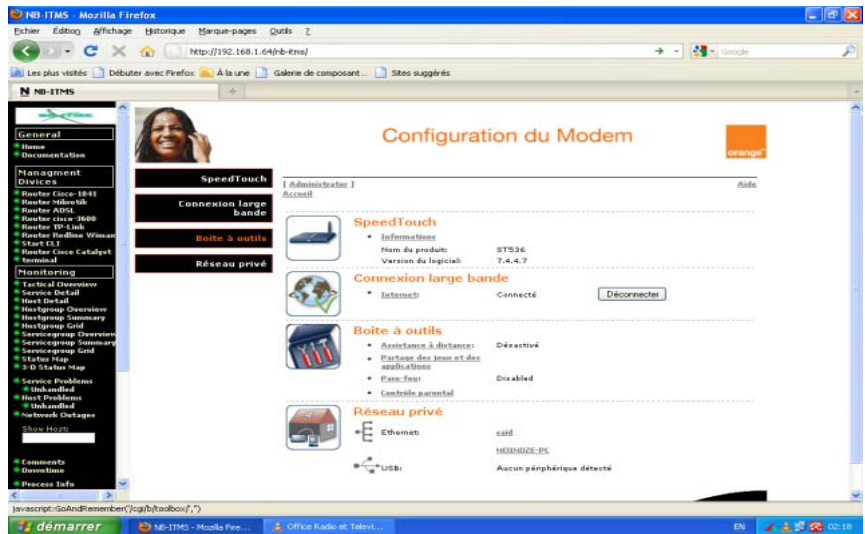

Figure 8b : Management configuration of Router ADSL Interface via NB-ITMS

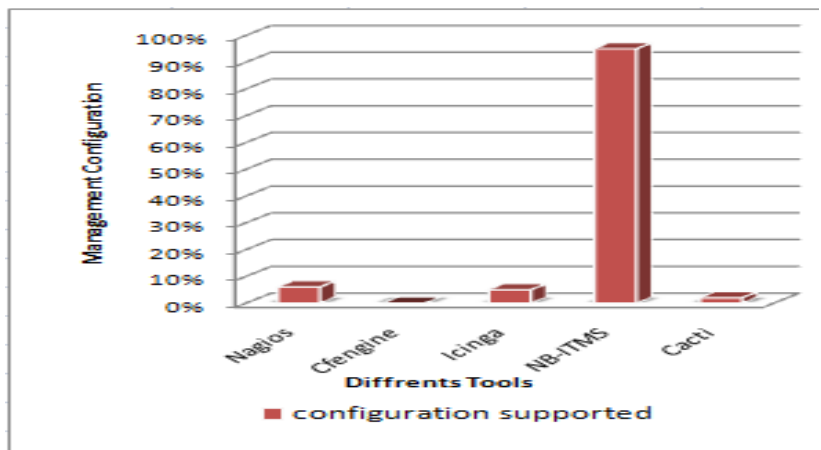

Figure 8 c : Configuration management

According to (1), we have allocated realistic values for each metric function and compared the results of simulations with four other tools. The tools that have been considered for comparison with NB-ITMS are Nagios, Cfengine, Icinga, MRTG and NB-ITMS. It is important to mention that we haven't considered in this work free tools running on a specific Linux distribution as EyesOfNetwork (EON), Full Automated Nagios (FAN) because of their limitation to a single distribution of Linux. For ''k'', any of the five functions of network management defined by ISO (International Standards Organization) and known by the acronym FCAPS (Fault, Configuration, Accounting Performance and Security) could be considered.

For simplicity purpose, we supposed that all these functions are equally important and each of the five management features are equally important so could be allocated a rate of 20%.

Thus, we have compared five tools in terms of fault and configuration management. Fig. 8a&b are examples of screen capture of NB-ITMS used respectively to supervise the MXC 9500 Alcatel-Lucent microwave station and configure operational modems of Orange Telco. Fig. 8c shows the results of our comparison. It can be noticed that almost all enumerated free tools are only providing monitoring management and do not provide configuration management. Nagios, Icinga, Cacti, are open source tools but they only support supervision without any possibility of configuration management. Cfengine is more appropriate for network and system administration.

## 4- Discussions on NB-ITMS efficiency

Fig. 8 shows that NB-ITMS ensures both fault detection of network elements (NE) and configuration management. According [David Padi, (2009)] for scoring and rating of the different solutions, we have obtained Figures 9.

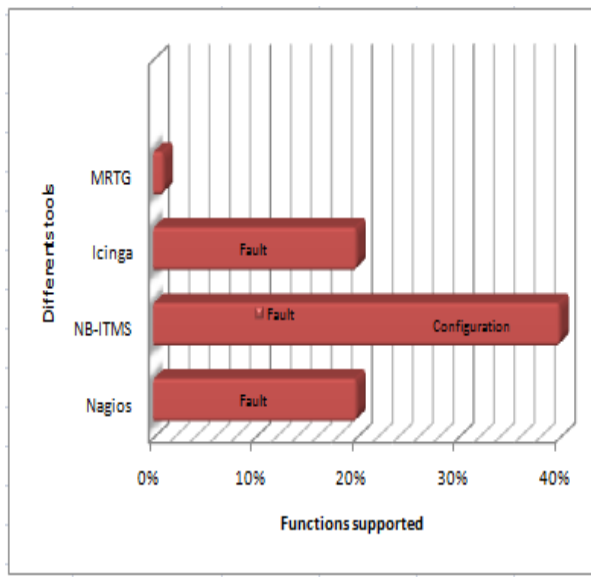 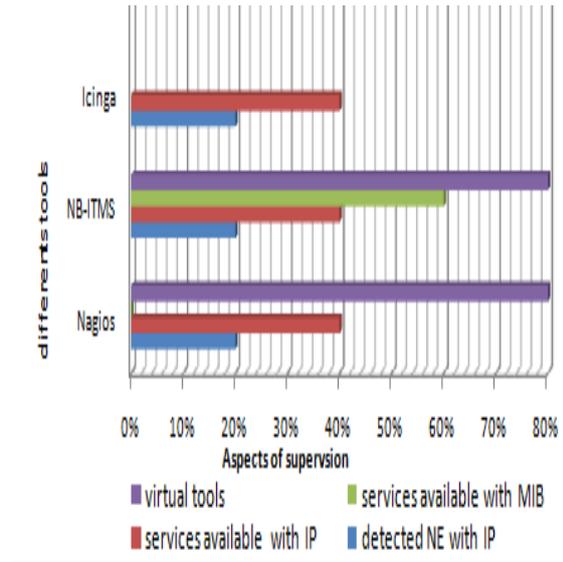

Figure 9: Integrated management functions            Figure 10: Aspects of supervision

Fig. 9 shows that NB-ITMS support fault and configuration management functions. Its rating according to the previous section 3 is 40% compared to the other open and free existing solutions.

We have considered detailed supervision aspects in Fig. 10. Compared to Nagios, NB-ITMS has integrated more MIBs of different manufacturers as in Fig 7a. Then for fault management function, Fig. 10 shows that NB-ITMS is slightly more efficient than Nagios. For faults management, if the different tools costs are considered as the same, the operational cost of these tools will be approximately the same, Fig. 8c&10 shows that the quality of service is globally better for NB-ITMS than Nagios, Cfengine or Icinga.

For management of configuration, Fig. 8 shows that NB-ITMS has also the best performance.
Then, we can conclude that NB-ITMS is more efficient than Nagios, Cfengine and Icinga for fault and configuration management functions.

5. Conclusion

Operators are facing a critical situation nowadays aiming at reducing operational costs of their infrastructure. For that, there are looking for low cost and efficient tools to manage their more and more complex infrastructure. Almost all existing open source are only providing monitoring features and no configuration function. In this work, we have proposed an extension of Nagios open tools to enable configuration functionality. We have proposed a framework to cope with the heterogeneity of underlying infrastructure and services. Hence, we have enhanced the GUI to enable the system to label the map view and help the administrator to clearly identify the equipments. The description of the system has shown the main additional bloc built in. From this study, we have shown thanks to different comparisons performed against other existing open source management tools that NB-ITMS is providing enhanced feature to remotely configure most parts of underlying equipments.

Acknowledgment

We would like to thanks professor Nazim Agoulmine, from the university of Evry Val d'Essonne for his valuable remarks and guidance in this work.